\documentstyle[prb,aps,graphicx,epsf,amssymb]{revtex}

\begin{document}

\title{The Effects of Finite Length on the Electronic Structure of Carbon 
Nanotubes}

\author{Alain Rochefort\cite{rochefor}}
\address{Centre de Recherche en Calcul Appliqu\'e (CERCA), 5160 boul. 
D\'ecarie, bureau 400, Montr\'eal, (Qu\'ebec) Canada H3X 2H9}
\author{Dennis R. Salahub\cite{salahub}}
\address{D\'epartement de Chimie, Universit\'e de Montr\'eal, C.P. 6128,
Succ. Centre-Ville, Montr\'eal, (Qu\'ebec) Canada H3C 3J7\\
Centre de Recherche en Calcul Appliqu\'e (CERCA), 5160 boul. 
D\'ecarie, bureau 400, Montr\'eal, (Qu\'ebec) Canada H3X 2H9}
\author{Phaedon Avouris\cite{avouris}}
\address{IBM Research Division, T.J. Watson Research Center, P.O. Box 218,
Yorktown Heights, NY 10598, USA}
\vskip0.8cm
\date{\today}

\maketitle

\vskip1.0cm
\begin{center}
{\large Abstract}
\end{center}

\begin{abstract} 
The electronic structure of finite-length armchair carbon nanotubes has been
studied using several ab-initio and semi-empirical quantum computational
techniques.  The additional confinement of the electrons along the tube axis
leads to the opening of a band-gap in short armchair tubes.  The value of the
band-gap decreases with increasing tube length, however, the decrease is not
monotonic but shows a well defined oscillation in short tubes.  This
oscillation can be explained in terms of periodic changes in the bonding
characteristics of the HOMO and LUMO orbitals of the tubes.  Finite size
graphene sheets are also found to have a finite band-gap, but no clear
oscillation is observed.  As the length of the tube increases the density of
states (DOS) spectrum evolves from that characteristic of a zero-dimensional
(0-D) system to that characteristic of a delocalized one-dimensional (1-D)
system.  This transformation appears to be complete already for tubes 5-10 nm
long.  The chemical stability of the nanotubes, expressed by the binding
energy of a carbon atom, increases in a similar manner.
\end{abstract}

\section{INTRODUCTION}

Carbon nanotubes are a new form of carbon with rather unique properties
\cite{iijima,yacobson,dresselhaus}. A single wall nanotube can be considered
as resulting from rolling up a single graphene sheet to form a hollow
cylinder.  The chirality and diameter of a particular tube can be described
in terms of a role-up vector $C=n\vec a + m\vec b \equiv (n,m)$, where $\vec
a$ and $\vec b$ denote the unit vectors of the hexagonal honeycomb lattice,
and n and m are integers \cite{dresselhaus}. In a 2-D graphene sheet the
$\pi$-bonding and $\pi^*$-antibonding states become degenerate at the K-point
of the hexagonal Brillouin zone, resulting in a zero band-gap semiconductor.
In a nanotube, quantization of the wavefunction along its circumference
restricts the allowed wave-vectors to certain directions of the graphite
Brillouin zone so that  $C \cdot k = 2 \pi j$,  where $j$ is an integer.  If
at least one of these wavevectors passes through the K point, the tube is
metallic, otherwise it is a semiconductor with a finite band-gap.  Thus,
$(n,0)$ zig-zag tubes are expected to be metallic if $n/3$ is an integer, and
semiconducting otherwise \cite{saito,mintmire,blase}. As the role-up vector
$C$ rotates away from the $(n,0)$ direction, the resulting $(n,m)$ tubes are
chiral and are expected to be metallic if $(2n+m)/3$ is an integer.
Otherwise they are semiconductors with a gap $\propto 1/R$, where $R$ is the
tube radius.  Finally, when $C$ is rotated 30$^{\circ}$  away from the
$(n,0)$ direction, $n=m$ and the resulting armchair tubes are metallic.
Recent scanning tunneling microscope studies have verified the basic
conclusions of simple theory \cite{wildoer,odom}.\\

The above predictions assume a semi-infinitely long tube, and corresponding
experimental electronic structure studies have utilized tubes several microns
in length.  It is very important to know how an additional confinement of the
electrons, along the nanotube axis, in finite length tubes will affect the
tube's electronic structure.  As the tube length is diminished, there should
be a transition from a 1-D structure (quantum wire) to a 0-D structure
(quantum dot).  Currently, little is known about the effects of finite length
on the properties of carbon nanotubes.  Recently, however, Venema {\it et
al.} \cite{venema} have been able to cut nanotubes into segments of a few
tens of nanometers length and, using tunneling spectroscopy, obtained
evidence of increased band-gaps in these short tubes.  Understanding quantum
size effects in carbon nanotubes, besides the basic scientific value of
understanding a prototypical 1-D to 0-D transition, is essential in device
applications of nanotubes. The miniaturization offered by nanotube-based
devices requires that not only their diameter but also their length should be
in the nanometer  range.  For example, micrometer-long nanotubes have been
used as Coulomb islands in single electron transistors (SET) operating at
liquid helium temperatures \cite{tans1,bockrath}.  By using shorter tubes,
the operating temperature of SETs can be raised drastically.  In general, the
electrical properties of nanotubes may be tailored by selecting their
length.\\

To obtain the needed insight into the electronic structure of short nanotubes
and learn about their stability (information useful in studies of growth
mechanisms) we performed electronic structure calculations on different
length $(6,6)$ armchair nanotubes.  To determine the local properties and the
spatially-resolved electronic structure of finite-size nanotubes,
quantum-chemical methods are more suitable than band-structure calculations.
However, the theoretical determination of electronic properties, such as the
band-gap value, remains a difficult task, as has been shown in studies of low
band-gap polymers \cite{roncali}. To compensate for the weaknesses of the
various quantum chemical techniques, we utilized a number of both
first-principles (density functional and Hartree-Fock) and semi-empirical
computational techniques (MNDO-PM3 and extended H\"uckel).  We used the above
techniques to calculate the band-gap values, the DOS spectrum and the binding
energy of carbon atoms as a function of the length of the tube segment.
While, according to the discussion in the Introduction, infinitely long
$(6,6)$ nanotubes should be metallic, all computational techniques predict a
band-gap for short nanotubes whose value decreases towards zero with
increasing tube length. Interestingly, the variation of the band-gap is not
monotonic but shows strong oscillations as a function of tube length.  While
all computational techniques used predict these oscillations, the actual
values of the band-gap vary strongly with the technique used.  Computed DOS
spectra and local density of states distributions show that the transition
from a 0-D to a 1-D-like spectrum takes place already in very short 5-10 nm
tubes.  The binding energy of the carbon atoms increases with increasing tube
length in a similar manner.  The behavior of finite graphene sheets is also
investigated.

\section{COMPUTATIONAL DETAILS}

Electronic structure calculations were performed using both ab-initio:
Hartree-Fock (HF) and Density Functional Theory (DFT), and semi empirical:
Modified Neglect of Differential Overlap - Parametrized Model 3 (MNDO-PM3)
and extended H\"uckel (EHMO) computational methods.  In the HF calculations,
we employed a parallel version of the GAMESS program \cite{gamess} in which
STO-3G quality basis sets were employed for carbon and hydrogen. For the DFT
computations, we used the deMon-KS software \cite{deMon1,deMon2,deMon3}. The
calculations were performed with Huzinaga's minimal basis sets
\cite{huzinaga} in conjunction with the generalized gradient approximation
(GGA) of Perdew and Wang for exchange \cite{perdew86} and correlation
\cite{perdew91}. The semi-empirical calculations were carried out with the
MNDO-PM3 method \cite{mndo,pm3} included in the GAMESS program \cite{gamess},
while for the  EHMO calculations we used the program included in the YAeHMOP
package \cite{yaehmop}.\\

In the carbon nanotube models, the C-C and C-H bond length were fixed at the
values observed in bulk graphite \cite{graphite} which  are 1.42 and 1.09
{\AA}, respectively.  All calculations were performed at fixed geometries
without optimization of the atomic structures. The elementary unit used,
called here a section, is defined as a single circular plane of carbon atoms
that are packed along the length of the nanotube (note: the distance between
two consecutive sections in the nanotube is 1.22 {\AA}). The dangling bonds
at the ends of the tube were saturated with hydrogen atoms.  The energy
band-gap is obtained as the difference between the LUMO and HOMO
energies, while the Fermi energy ($E_F$ = 0 eV) is the mean energy between
LUMO and HOMO.\\

Density of states (DOS) plots were generated by convoluting the computed
electronic structure with a 50:50 combination of Gaussian and Lorentzian
functions.  In order to analyze the nature of the energy bands, we performed
a series of projections to obtain the local density of states (LDOS), where
each molecular orbital was weighted by the contribution obtained from a
Mulliken analysis of specific carbon atoms. The general expression used to
generate the DOS plots is :\\

\begin{displaymath} 
\textrm{DOS} (\epsilon) = 
\sum_{i=1}^{N} \bigg ( [
\frac{n_i}{{\omega} {(\frac{\pi}{2})}^{1/2}} \cdot \textrm{exp} { (\frac{-2
({\epsilon}- {\epsilon_i})^2}{{\omega}^2})} ] + [ \frac{2
{n_i}{\omega}}{{\pi} ({\omega}^2 + 4({\epsilon} - {\epsilon}_i)^2)}]\bigg)
\end{displaymath}

\noindent
where $\omega$ is the resolution, {\it n$_i$} is the population of the level
{\it i} and ($\epsilon$ - $\epsilon_{i}$) is the energy difference between
the energy $\epsilon$ and the eigenvalue $\epsilon_{i}$.

\section{RESULTS and DISCUSSION}

To evaluate the influence of finite length on the electronic properties of
the nanotubes, we calculated the electronic structure of different length
segments of a (6,6) nanotube.  As Figure 1 shows, all computational
techniques used predict that, unlike the infinite tube that is metallic, very
short ($<$ 100 {\AA}) nanotubes have an energy band-gap ($E_{g}$).
Furthermore, Figure 1 reveals that $E_{g}$ shows a regular oscillatory
dependence on tube length.  The oscillation of the band-gap values is
particularly strong in the HF method where $E_{g}$ varies from 2.8 to 9.0 eV
for tubes shorter than 20 {\AA}. The semi-empirical MNDO-PM3 method gives
results comparable to those obtained by HF.  On the other hand, extended
H\"uckel molecular orbital (EHMO) and density functional theory - generalized
gradient approximation (DFT-GGA) methods lead to quite similar values of the
band-gap.  The oscillation amplitude decreases, and the band-gap value
converges slowly to zero for larger nanotubes.  For a 96 {\AA} long nanotube,
EHMO predicts a metallic behavior while MNDO-PM3 still shows a band-gap of
2.0 eV.  The comparison of the different quantum-chemical methods used to
determine $E_{g}$ will be discussed in more detail below.\\

The inset in Figure 1 gives an expanded view of the band-gap oscillations for
nanotubes shorter than 25 {\AA}. All computational methods predict a regular
and converging oscillation of $E_{g}$.  The oscillation is related to the
structure of the nanotube; high band-gap values are invariably obtained for
tubes having $(3n+1)$ (where $n$ = 1,2,...) sections (reminder: a section
contains a single circular plane of carbon atoms).  Furthermore, consecutive
medium or low band-gap values are periodically observed for nanotubes having
$(3n)$ and $(3n-1)$ sections.  The main difference between the oscillatory
behavior of $E_{g}$ obtained with the different computational methods lies in
the different length needed for an inversion between medium and low band-gap
behavior to occur.  The different oscillations observed, as well as the low
to medium band-gap inversion for small nanotubes can be explained in terms of
the changing bonding characteristics of the HOMO and LUMO orbitals.\\

Figure 2 shows the nodal structure of the frontier $\pi$-orbitals for the
nanotube models, while Figure 3 gives the energies of ten molecular orbitals
(obtained from EHMO theory) which lie near the band-gap region.  The relative
stability of frontier orbitals (HOMO and LUMO) for the first three models
(S2, S3 and S4 with respectively, 2,3 and 4 sections) is based on the nature
of the bonding along the circumference of the tube, i.e intra-section
interactions, as well as the bonding along the tube axis, i.e.  inter-section
interactions. Furthermore, the S2, S3 and S4 models constitute the repeat
units from which the HOMO/LUMO orbitals of the longer tubes may be
constructed. From Figure 2, we see that the HOMO for S2 (two section model),
the first member of the S$(3n-1)$ group, has bonding character along the
circumference, while the LUMO has bonding character between sections, i.e.,
along the tube axis.  We should then expect the energy difference between
HOMO and LUMO of the S2 structure to be small, leading to a low band-gap.  On
going from the S3 (S$(3n)$ group) to the S4 (S$(3n+1)$ group) structures, we
observe from Figure 2 that the HOMO acquires an additional bonding character
along the tube axis and hence is stabilized more.  On the other hand, the
LUMOs of both S3 and S4 type structures show antibonding inter-section
interactions which lead to a larger band-gap. The gap is lower for S3 due to
the net bonding character of the LUMO along the circumference. The structure
and the stability of the frontier orbitals for longer nanotubes are directly
related to the characteristics of those three first structures. The S$(3n-1)$
group (i.e. S2, S5, S8, \ldots) is characterized by non-bonding inter-section
interactions, as is evident by considering the orbital nodal properties in
Figure 2.  The stability of the frontier orbitals in the S$(3n-1)$ group is
weakly influenced by inter-section interactions.  The slight destabilization
of the frontier orbitals is due to long-range antibonding lateral
interactions that reduce slowly the band-gap with increasing tube length.  In
the S$(3n)$ group (S3, S6, S9, \ldots), the decreasing band-gap is caused by
a destabilizing lateral antibonding interaction in the HOMO coupled with a
stabilizing lateral interaction in the LUMO.  The lateral interactions have a
strong influence on the value of the band-gap of the S$(3n)$ group so that it
actually becomes lower than the band-gap value of the S$(3n-1)$ group at
$n$=3.  This behavior occurs at different tube lengths depending on the
computational method used to determine the energies of the molecular
orbitals.  Finally, the S$(3n+1)$ group that is characterized by high
band-gaps follows a similar trend as the S$(3n)$ group with increasing
nanotube length; a destabilization of the HOMO through antibonding
interactions between sections and stabilization of the LUMO through partly
bonding lateral interactions.\\

Band-gap oscillations were also reported recently by Yoshizawa {\it et al.}
\cite{yoshizawa} for 2-D polyphenanthrenes on the basis of extended H\"uckel
band structure calculations.  Since, instead of the carbon section used here,
a phenanthrene-edge structure was used as the elementary building unit in
that study, high band-gaps were found in systems with $(6n-2)$ sections
rather than $(3n+1)$ sections.  Nevertheless, although the compositions of
the frontier orbitals of the $(6,6)$ armchair nanotube are to a certain
extent different from those of the  periodic 2-D polyphenanthrenes, both
structures give rise to a similar pattern of band-gap oscillations.  In
Figure 4 we compare the variation of the band-gap value for a perfect $(6,6)$
nanotube to that of an equivalent 2-D graphene fragment (whose broken C-C
bonds were saturated with hydrogen atoms), and that of the periodic sheet of
2-D polyphenanthrenes as determined by Yoshizawa \cite{yoshizawa}.  We find
that the growth of the finite 2-D graphene sheet does not lead to a clear
band-gap oscillation pattern and the $E_{g}$ rapidly converges to low
values.  Furthermore, the finite graphene sheet gives quite different $E_{g}$
values than the infinite polyphenanthrene sheet.  Figure 5 shows the
influence of the nanotube diameter on the magnitude of the band-gap
oscillations with respect to the fluctuations observed for a periodic
graphene sheet.  Increasing the nanotube diameter by going from the $(6,6)$
to the $(10,10)$ tube, brings its band-gap value closer to that of the
periodic 2-D graphene sheet.  However, the 1/(diameter) scaling of $E_{g}$
predicted for infinite tubes \cite{dresselhaus,mintmire} is not observed in
very short tube segments.\\

Crucial to the understanding of the nanotube growth process are changes in
the binding energy of C atoms as a function of the increasing nanotube
length.  Figure 6 compares the binding energy (BE) of carbon atoms in a
finite (6,6) nanotube as determined by different computational techniques.  A
converged BE value indicates that all carbon atoms are nearly equivalent to
each other.  Major changes are observed only for very short tubes where the
BE found to increase sharply up to about 15 {\AA}, and then changes more
gradually upto 96 {\AA} (not shown).  The smaller changes in BE observed with
the EHMO method can be understood as due to the treatment of long-range
interaction in EHMO theory; carbon atoms are weakly perturbed by second and
higher neighbor carbon atoms.  The highest estimate of BE is obtained with
MNDO-PM3 (7.0 eV / C atom), and the lowest with EHMO (5.5 eV / C atom).  The
BE obtained with the ab-initio DFT-GGA (estimated) and HF lie between these
two extremes.  The low binding energy obtained for short tubes suggests that
carbon in short nanotubes would be more reactive.\\

Although the energetic properties of nanotubes seem to converge rapidly, a
sufficiently long nanotube model is necessary to reproduce the main
characteristics of the electronic structure of a quasi-1D periodic system.
The influence of the length on the general electronic structure of a $(6,6)$
armchair nanotube is reported in Figure 7 obtained using the EHMO method.
The nanotube electronic structure is characterized in terms of the position
and the intensity of the main DOS peaks, the magnitude of DOS at the Fermi
level, and the presence of fine structure.  The spectrum of the very short 4
{\AA} (4 carbon sections) model is characteristic of a 0-D (quantum dot)
system with discrete quantum levels.  However, the band position and its
overall DOS profile is not drastically different from those of more extended
systems. The DOS profile for a 96 {\AA} long model, on the other hand,
clearly reveals the presence of characteristic fine structures at high
binding energy (BE) between -22 and -12 eV. These structures are ``van Hove
singularities" characteristic of periodical one-dimensional (1-D) systems
\cite{ashcroft}.  Each of these peaks is characterized by a specific number
of nodes in the wavefunction along the circumference of a single nanotube
section, while the $1/{\sqrt{E-E_0}}$ tail reflects the free electron
character along the tube axis.  In fact, these singularities are becoming
evident even in the 47 {\AA} nanotube that represents the minimum length
nanotube that exemplifies such 1-D-like characteristics. For nanotubes larger
than 47 {\AA}, the main changes in electronic structure occur around the
Fermi level, where the DOS is found to increase slowly with increasing
nanotube length.  Although it gives a different estimate for the band-gap,
the semi-empirical MNDO-PM3 method shows similar trends as EHMO concerning
the electronic structure of the growing nanotubes.\\

Figure 8 gives a more detailed description of the electronic
structure of the 96 {\AA} nanotube \cite{rochefort}. In the general DOS
diagram, the van Hove singularities at high BE are $\sigma$-states that
originate from C$(2s)$ atomic orbital combinations.  The $\sigma$-states
formed from C$(2p)$ orbitals lie at lower BE between -11 and -4 eV, while
$\pi$-states extend from -5 to 0 eV.  These states are important as they are
responsible for the electrical properties of nanotubes.  The states above the
Fermi level are mainly ${\pi}^*$-states, while ${\sigma}^*$-states extend
above 10 eV.  The inset gives an expanded view of the fine structure of the
96 {\AA} nanotube near the Fermi energy.  In addition, the values of the
local density of states (LDOS) near the Fermi energy for sections ranging
from the tube boundary (index 1) to the middle of the tube (section 40) are
plotted.  To construct these LDOS plots, we summed over the density of the
twelve carbon atoms contained in that section of the nanotube.  From figure 8
it can be seen that the LDOS of the ${\pi}$-states near $E_F$ hardly
changes along the length of the tube, indicating that these states are indeed
well delocalized within this short tube.\\

Finally, we shall compare briefly the results obtained with the different
computational techniques. As shown in Figure 1, both HF and MNDO-PM3 methods
give a large band-gap (9.0 and 6.5 eV, respectively) for a short 4 {\AA}
nanotube, while DFT and EHMO give moderate band-gap values (2.4 and 1.8 eV,
respectively). Due to the poor description of unoccupied orbital energy
levels in Hartree-Fock theory, the band-gap values are overestimated by
several electron volts for low band-gap polymers
\cite{hunt,kertesz,salzner}.  The problem is slightly less important in the
MNDO-PM3 method, where the parametrization includes a treatment of
correlation through the fitting with experimental data
\cite{bakowies1,dewar,thiel}. Previous MNDO-band structure calculations have
predicted a metallic behavior ($E_{g}$ = 0 eV) for periodic 2-D
graphite\cite{lee,bakowies2}. This suggests that the computed band-gap should
converge to zero for extended nanotube models; we estimate from Figure 1 that
nanotubes larger than ~450 {\AA} would be metallic. On the other hand,
DFT-based methods tend to systematically underestimate the band-gap value of
low band-gap polymers \cite{dreizler}.  However, recent developments of
DFT/hybrid functionals that include a weighted contribution of Hartree-Fock
and DFT exchange allow a better evaluation of the band-gap value
\cite{salzner,salzner2}.  Given that the computed band-gaps with EHMO are
similar to the DFT results, we may conclude that the band-gap values are also
underestimated with EHMO.  Based on DFT and EHMO results, nanotubes longer
than 100 {\AA} would be metallic.  However, given the above discussion on the
limitations of the computational techniques, the transition to the metallic
state likely occurs at longer lengths, probably in the range of 10-20 nm.

\section{CONCLUSIONS}

We have investigated the electronic structure of finite-length armchair
carbon nanotubes using a number of ab-initio and semi-empirical quantum
chemistry techniques.  Electrons in carbon nanotubes are confined along their
circumference.  In this study we imposed an additional confinement along the
tube axis by using short tube segments.  We found that armchair tubes that
are metallic when infinitely long, develop a band-gap.  The value of the
band-gap decreases with increasing tube length, but the decrease is not
monotonic; it shows a well defined oscillation in short tubes.  This
oscillation can be accounted for in terms of the changes in the bonding
characteristics of the HOMO and LUMO orbitals of the tubes as a function of
their increasing length.  Finite size graphene sheets were also found to have
a finite band-gap, but no clear oscillation was observed.  The DOS spectrum
of short nanotubes evolves with increasing length from that characteristic of
a 0-D system to that characteristic of a delocalized 1-D system.  This
transformation is complete already for tubes about 10 nm long.  The chemical
stability of the nanotubes, expressed by the binding energy of a carbon atom,
is low for short tubes but increases and saturates at tube lengths $<$ 10
nm.

\newpage

\begin{figure}
\caption{Variation of the band-gap of a $(6,6)$ nanotube 
as a function of its length determined using different computational
techniques. The inset gives an expanded view of the band-gap behavior
of very short tubes.}
\label{Figure 1.}
\end{figure}

\begin{figure}
\caption{Qualitative description of HOMO and LUMO orbitals as a function of
the nanotube length. The black and white circles indicate opposite phases
of the molecular wavefunction.}
\label{Figure 2.}
\end{figure}

\begin{figure}
\caption{Molecular orbital diagram for ten orbitals with energies near
the band-gap as determined using EHMO.}
\label{Figure 3.}
\end{figure}

\begin{figure}
\caption{Comparison of the band-gap variation for finite and periodic
graphene sheet with respect to the $(6,6)$ nanotube structure. The ``finite"
sheet is the unwrapped structure of the $(6,6)$ nanotube where broken
C-C bonds were saturated with hydrogen.}
\label{Figure 4.}
\end{figure}

\begin{figure}
\caption{Influence of the nanotube diameter on the band-gap oscillations
and comparison with the results obtained by Yoshizawa {\it et al.} 
\cite{yoshizawa} for an equivalent periodic 2D graphene sheet .}
\label{Figure 5.}
\end{figure}

\begin{figure}
\caption{Variation of the carbon atom binding energy as a function
of the length of the growing $(6,6)$ nanotube.}
\label{Figure 6.}
\end{figure}

\begin{figure}
\caption{Electronic structure of a growing (6,6) carbon nanotube obtained 
using the EHMO technique. Energy resolution ($\omega$) = 0.2 eV}
\label{Figure 7.}
\end{figure}

\begin{figure}
\caption{Total (DOS) and local density of states (LDOS) diagrams of a 
96 {\AA} armchair $(6,6)$ nanotube (resolution = 0.2 eV). The indices
in the LDOS diagram give the relative position of the carbon atoms in 
the nanotube structure (1: boundary, 40: middle of the nanotube)
The inset gives an expanded view of the DOS near the Fermi level (E=0 eV). 
The zero of the DOS scale is indicated by the
horizontal line, and the energy resolution is 0.05 eV.}
\label{Figure 8.}
\end{figure}

\clearpage
\newpage

\begin{figure}[p] 
\begin{center}
\includegraphics[width=17cm]{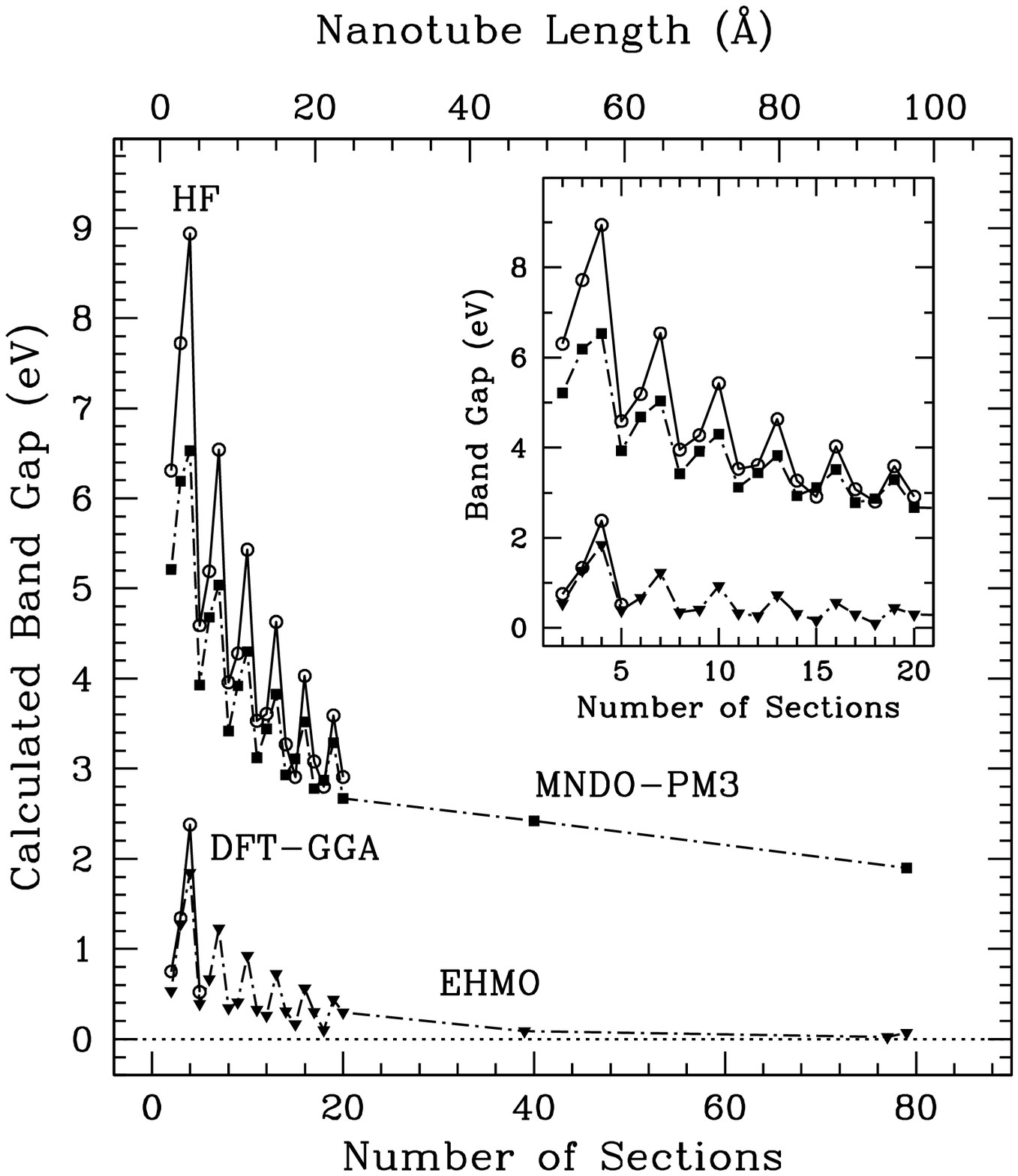} 
\end{center} 
Figure 1.\\
\end{figure}

\clearpage
\newpage

\begin{figure}[p] 
\begin{center}
\includegraphics[width=14cm]{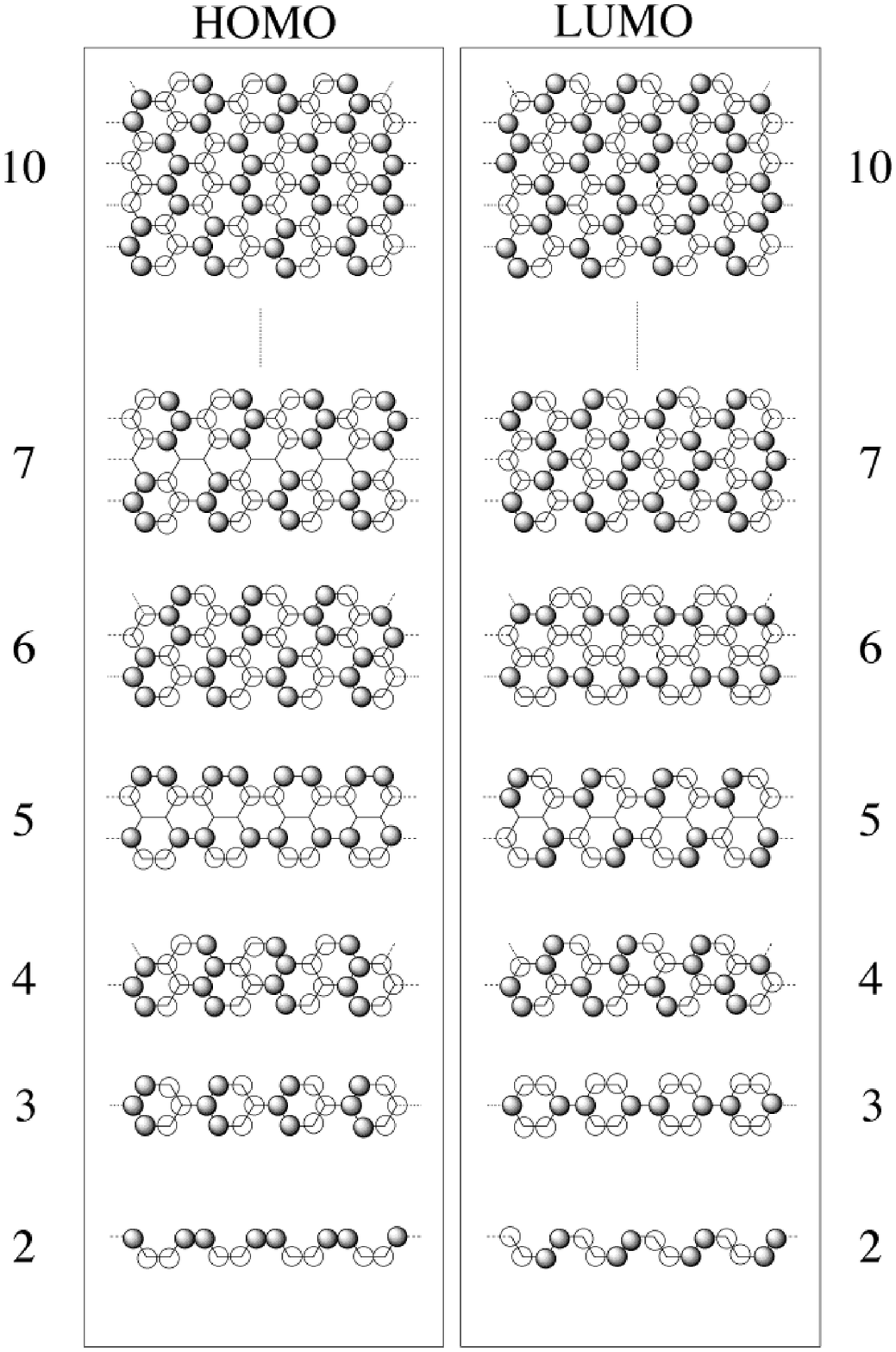} 
\end{center} 
Figure 2.\\
\end{figure}

\clearpage
\newpage

\begin{figure}[p] 
\begin{center}
\includegraphics[width=17cm]{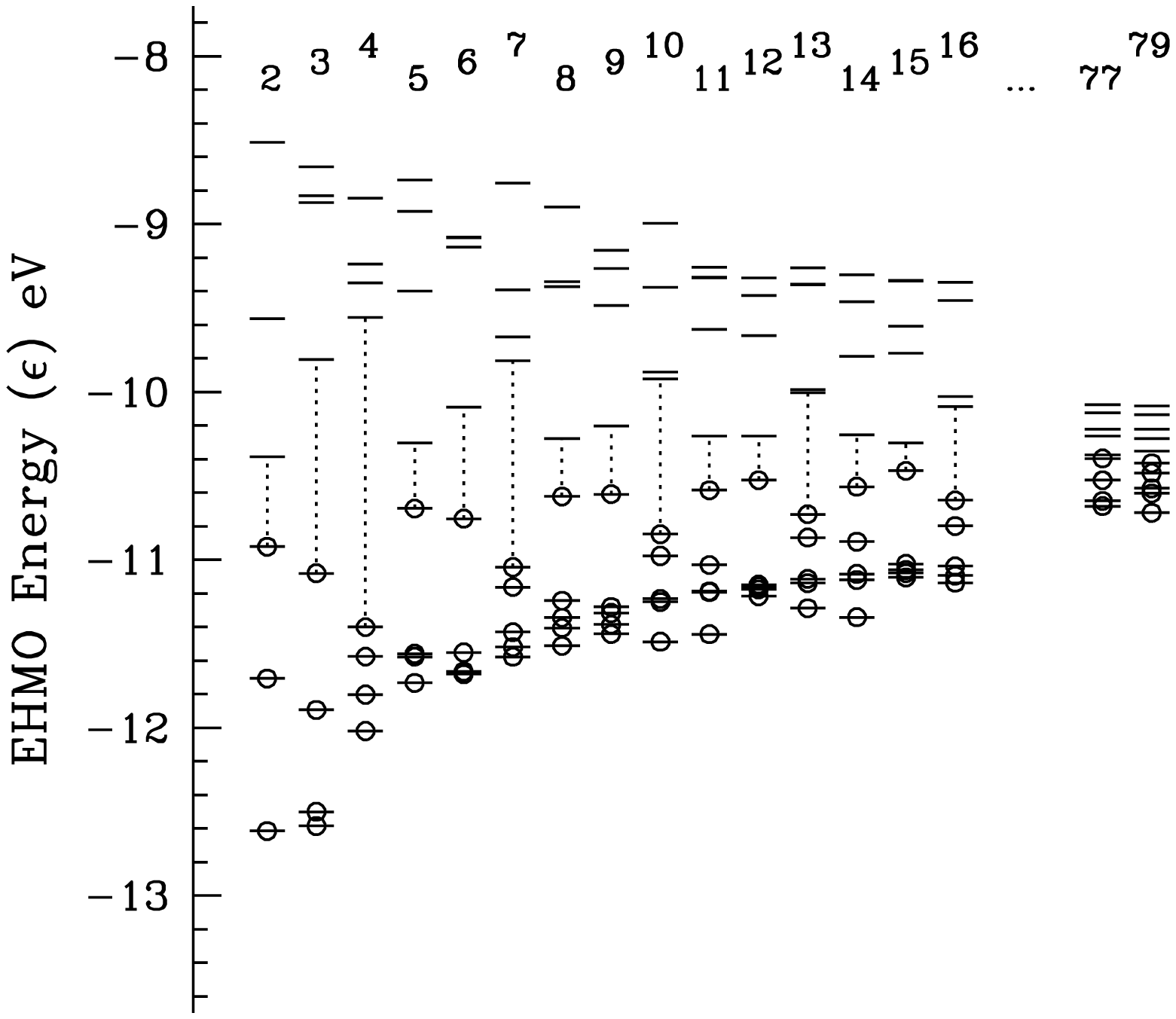} 
\end{center} 
Figure 3.\\
\end{figure}

\clearpage
\newpage

\begin{figure}[p] 
\begin{center}
\includegraphics[width=17cm]{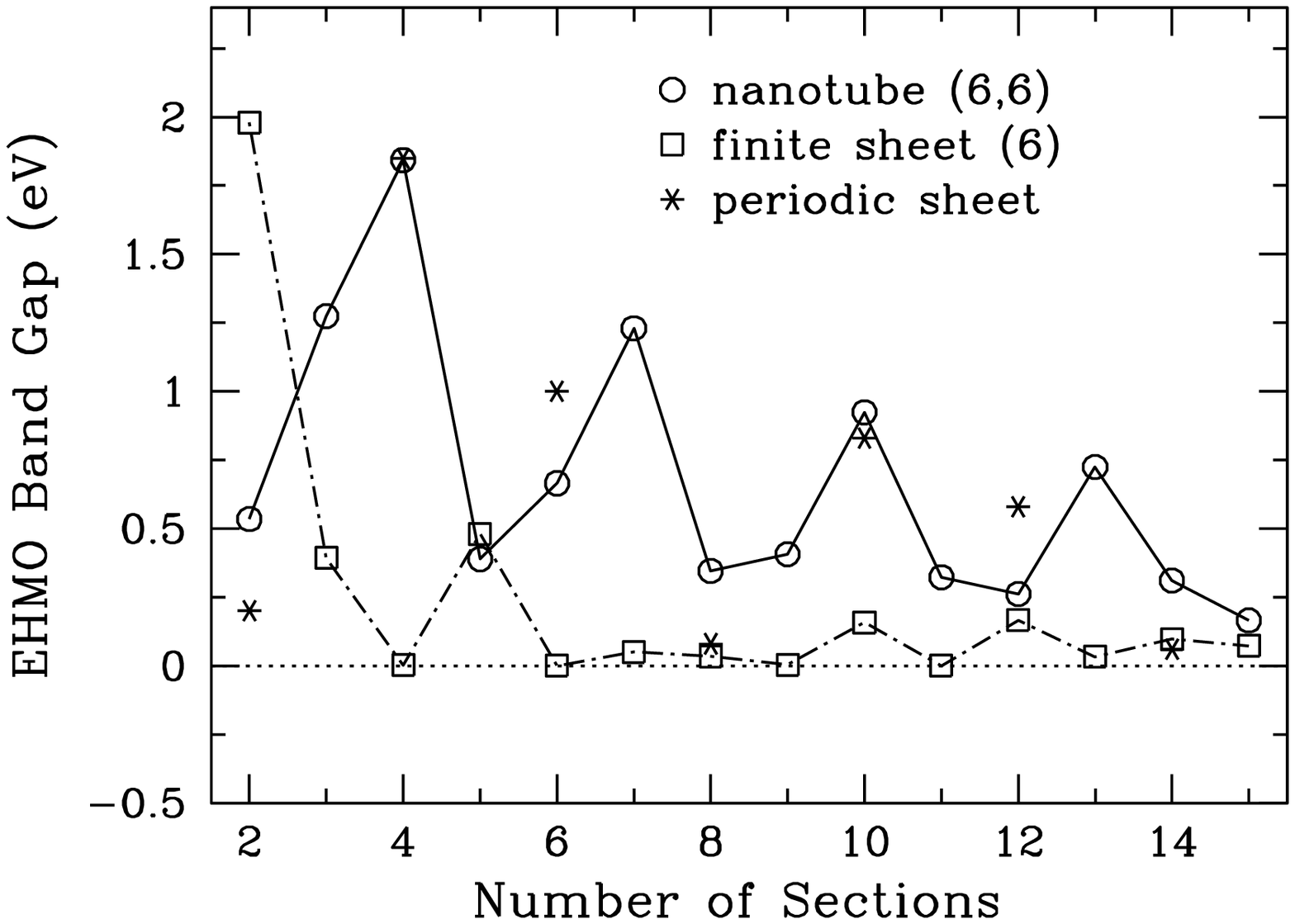} 
\end{center} 
Figure 4.\\
\end{figure}

\clearpage
\newpage

\begin{figure}[p] 
\begin{center}
\includegraphics[width=17cm]{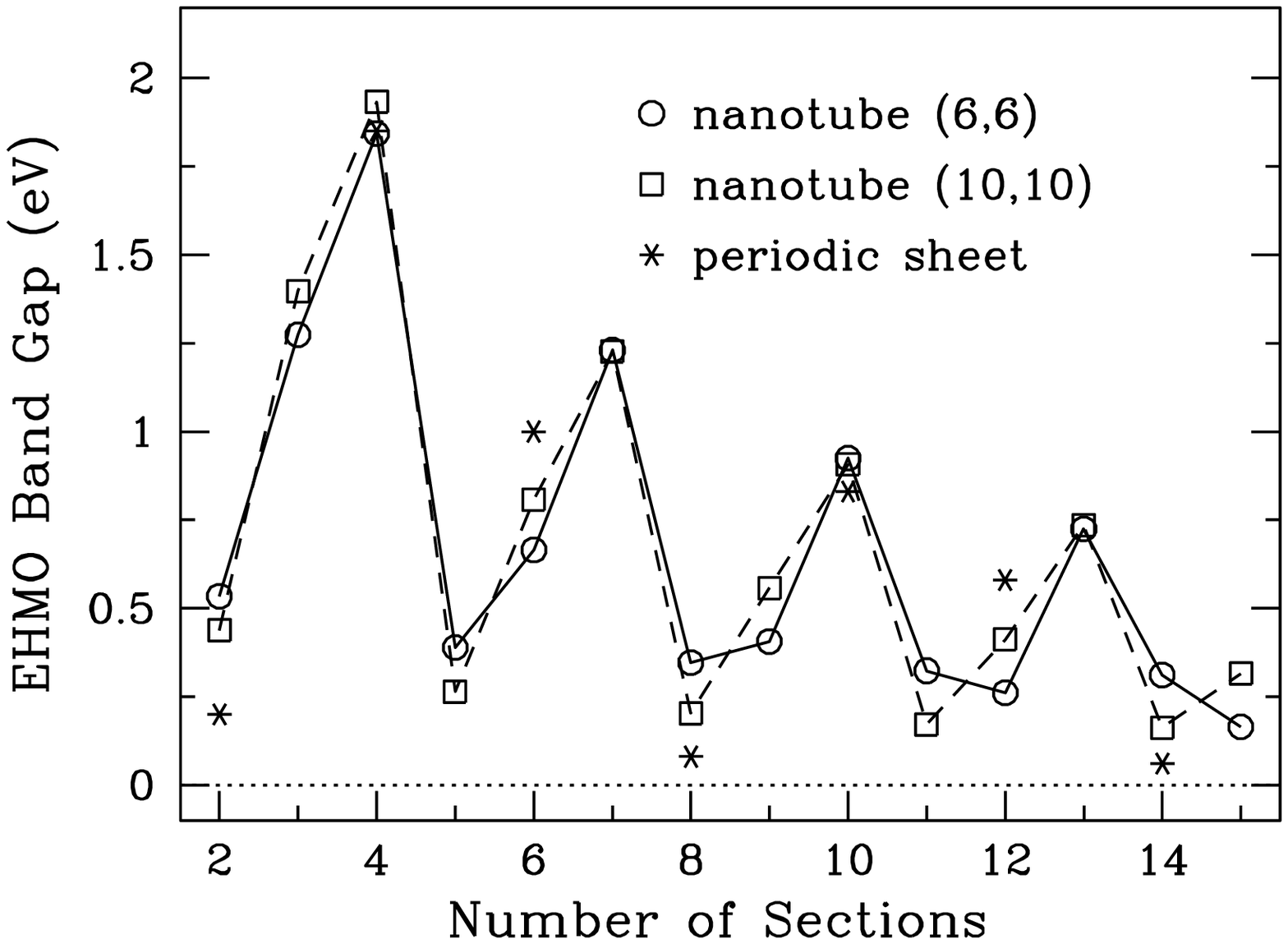} 
\end{center} 
Figure 5.\\
\end{figure}

\clearpage
\newpage

\begin{figure}[p] 
\begin{center}
\includegraphics[width=17cm]{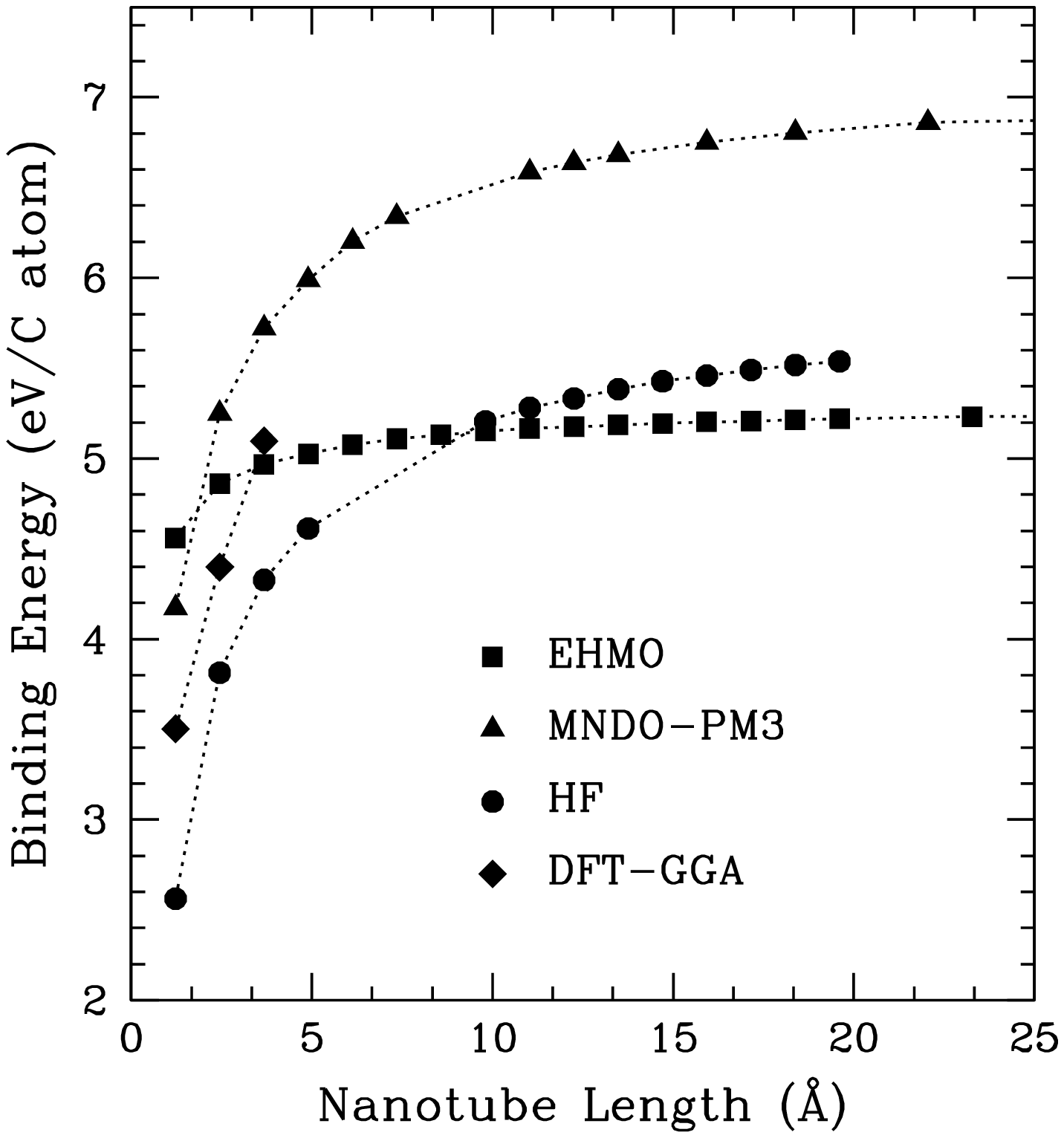} 
\end{center} 
Figure 6.\\
\end{figure}

\clearpage
\newpage

\begin{figure}[p] 
\begin{center}
\includegraphics[width=17cm]{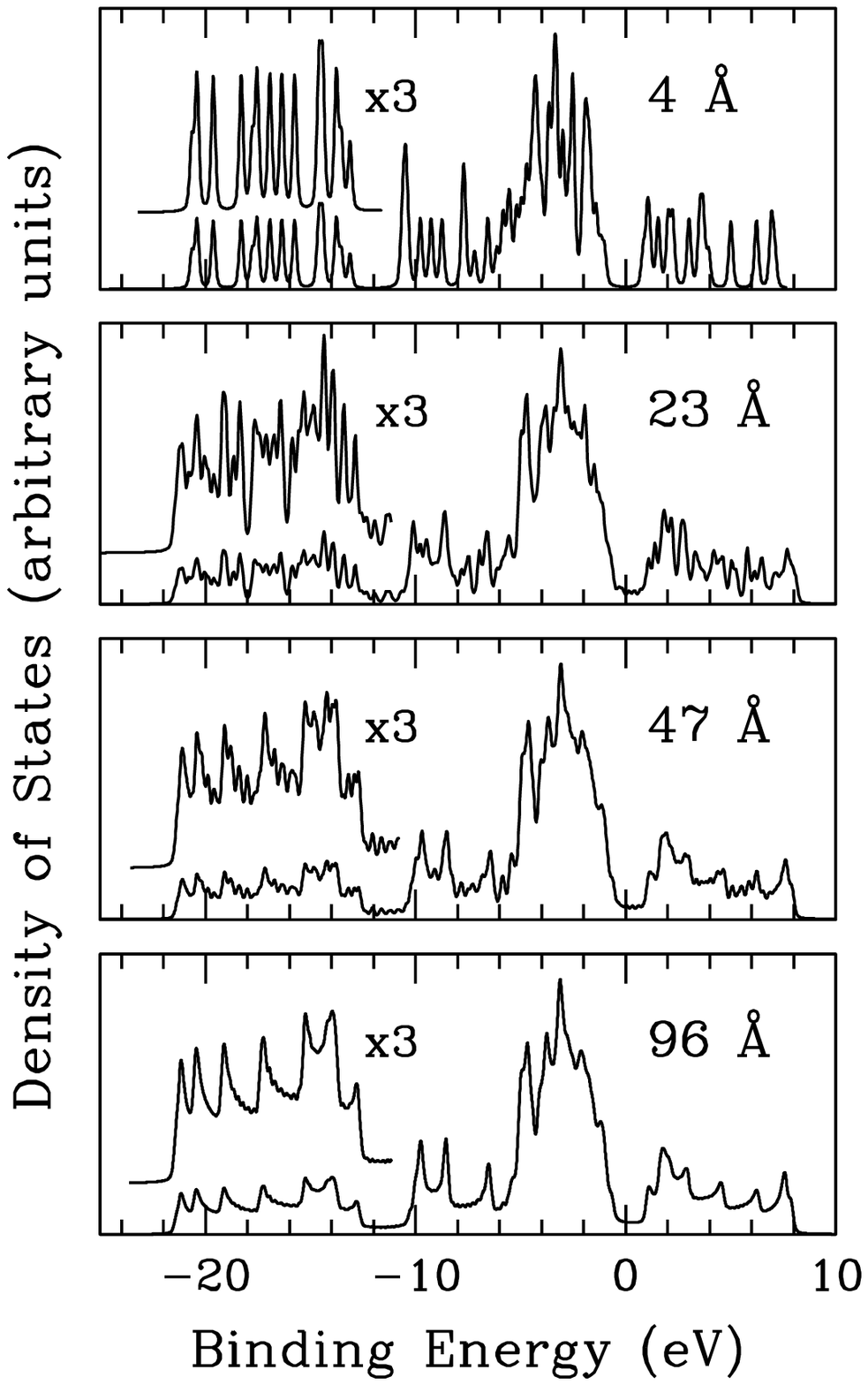} 
\end{center} 
Figure 7.\\
\end{figure}

\clearpage
\newpage

\begin{figure}[p] 
\begin{center}
\includegraphics[width=17cm]{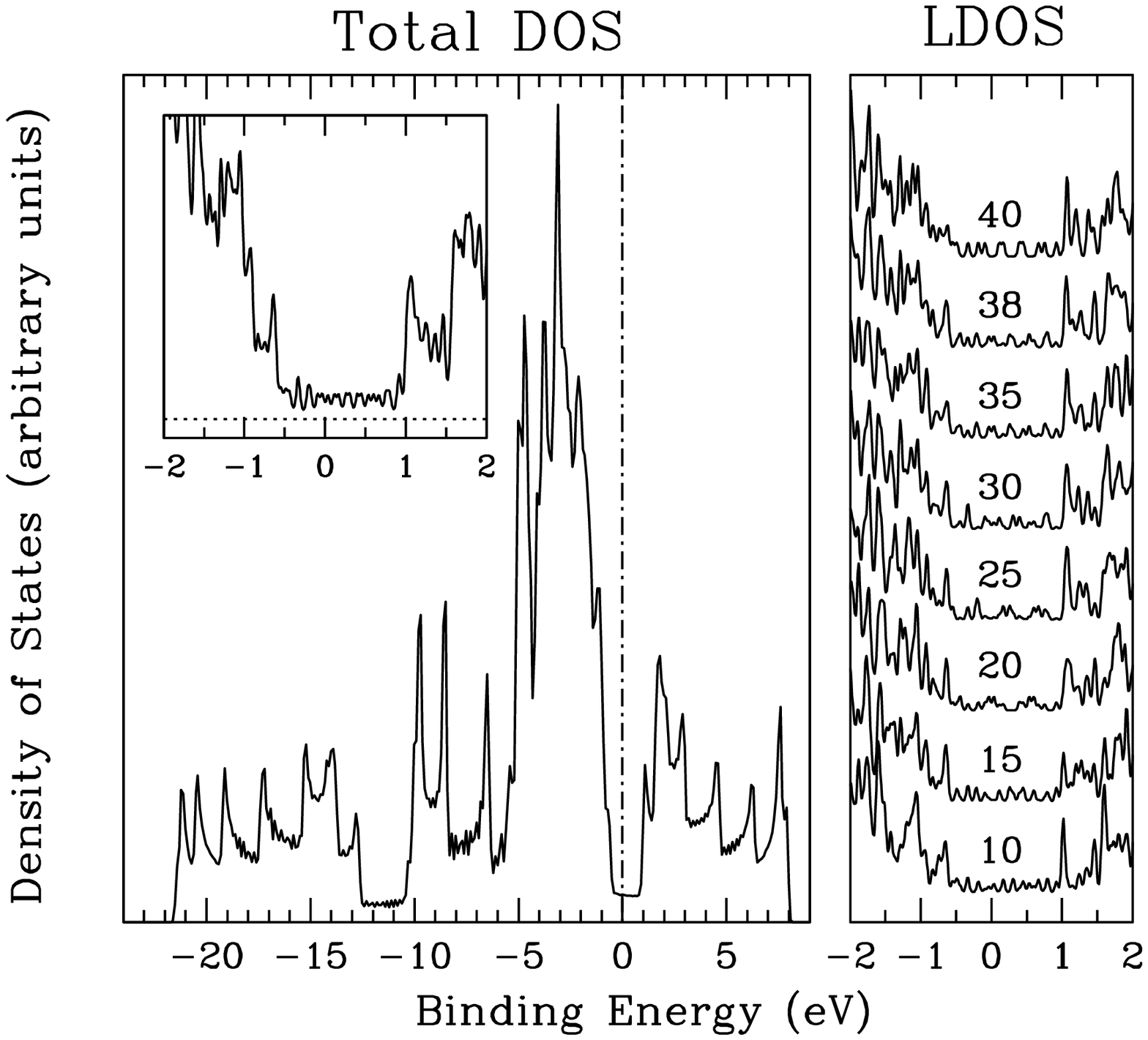} 
\end{center} 
Figure 8.\\
\end{figure}


\begin{references}
\bibitem[*]{rochefor} e-mail: rochefor@cerca.umontreal.ca\
\bibitem[\dagger]{salahub} e-mail: Dennis.Salahub@umontreal.ca\
\bibitem[\ddagger]{avouris} e-mail: avouris@us.ibm.com\
\bibitem{iijima} S. Iijima, {\it Nature}, {\bf 354} (1991) 56.\
\bibitem{yacobson} B.I.Yacobson and R.E. Smalley, {\it Am. Sci.} {\bf 85}
(1997) 324.\
\bibitem{dresselhaus} M.S. Dresselhaus, G. Dresselhaus and P.C. Eklund, 
{\it Science of Fullerenes and Carbon Nanotubes} (Academic Press, 
San Diego, 1996).\
\bibitem{saito} R. Saito, M. Fujita, G. Dresselhaus and M.S. Dresselhaus, 
{\it Appl. Phys. Lett.} {\bf 60} (1992) 2204.\
\bibitem{mintmire} J.W. Mintmire, B.I. Dunlap and C.T. White, {\it Phys. Rev.
Lett.} {\bf 68} (1992) 631.\
\bibitem{blase} X. Blase, L.X.  Benedict, E.L. Shirley and S.G. Louie 
{\it Phys. Rev. Lett.} {\bf 72} (1994) 1878.\
\bibitem{wildoer} J.W.G. Wild\"oer, L.C. Venema, A.G. Rinzler, R.E. Smalley 
and C. Dekker, {\it Nature} {\bf 391} (1998) 59.\ 
\bibitem{odom} T.W. Odom, J.-L. Huang, P. Kim and C.M. Lieber, {\it Nature} 
{\bf 391} (1998) 62.\
\bibitem{venema} L.C. Venema, J.W.G. Wild\"oer, H.L.J. Temminck Tuinstra, C. 
Dekker, A.G. Rinzler and R.E. Smalley, {\it Appl. Phys. Lett.} {\bf 71} 
(1997) 2629.\
\bibitem{tans1} S.J. Tans, M.H. Devoret and C. Dekker, {\it Nature} {\bf 386} 
(1997) 474.\
\bibitem{bockrath} M. Bockrath, D.H. Cobden and R.E. Smalley, {\it Science} 
{\bf 275} (1997) 1922.\  
\bibitem{roncali} J. Roncali, {\it Chem. Rev.} {\bf 97} (1997) 173.\
\bibitem{gamess} M.W. Schmidt, K.K. Baldridge, J.A. Boatz, S.T. Elbert, 
M.S. Gordon, J.H. Jensen, S. Koseki, N. Matsunaga, K.A. Nguyen, S.J. Su, 
T.L. Windus, M. Dupuis and J.A. Montgomery  {\it J. Comput. Chem.} {\bf 14} 
(1993) 1347-1363.\
\bibitem{deMon1} A. St-Amant and Salahub, D.R.  {\it Chem.Phys.Lett.} 
{\bf 169} (1990) 387.\ 
\bibitem{deMon2} A. St-Amant, {\it Ph.D. Thesis}, Universit\'e de Montr\'eal, 
1992.\ 
\bibitem{deMon3} deMon-KS version 3.4, M.E. Casida, C. Daul, A. Goursot, 
A. Koester, L. Pettersson, E. Proynov, A. St-Amant, D.R. Salahub, H. Duarte,
N. Godbout, J. Guan, C. Jamorski, M. Leboeuf, V. Malkin, O. Malkina,
F. Sim, and A. Vela, deMon Software, 1996.\ 
\bibitem{huzinaga} S. Huzinaga and J. Andzelm, {\it Gaussian basis sets for 
molecular calculations}, Physical sciences data; Vol.16, Amsterdam ; 
New York : Elsevier, 1984.\
\bibitem{perdew86} J.P. Perdew and Y. Wang, {\it Phys. Rev. B} {\bf 33} 
(1986) 8800.\
\bibitem{perdew91} J.P. Perdew and Y. Wang, {\it Phys.Rev. B} {\bf 46} 
(1992) 12947.\ 
\bibitem{mndo} M.J.S. Dewar and W. Thiel {\it J. Am. Chem. Soc.} {\bf 99} 
(1977) 4899-4907.
\bibitem{pm3} J.J.P. Stewart {\it J. Comput. Chem.} {\bf 10} (1989) 209-220.\
\bibitem{yaehmop} G. Landrum, {\it YAeHMOP} (Yet Another Extended
H\"uckel Molecular Orbital Package, Cornell University, Ithaca,
NY, 1995).\
\bibitem{graphite} Gmelin, {\it Handbuch der Anorganishen Chemie}, 8th ed.,
Verlag Chemie, Weinheim, 1968, Vol.14B/2, p.413. \
\bibitem{yoshizawa} K. Yoshizawa, K. Yahara, K. Tanaka and T. Yamabe,
{\it J. Phys. Chem.} {\bf 102} (1998) 498.\
\bibitem{ashcroft} N.W. Ashcroft and N.D. Mermin, {\it Solid States Physics}
Saunders College Publishing, Philadelphia, (1976).\
\bibitem{rochefort} The EHMO results obtained for the 96 {\AA} nanotube 
model are included in a previous publication submitted to {\it Chem. Phys. Lett.}
by the authors.\
\bibitem{hunt} W.J. Hunt and W.A. Goddard III, {\it Chem. Phys. Lett.}
{\bf 3} (1969) 414.\
\bibitem{kertesz} M. Kert\'esz, {\it Adv. Quantum Chem.} {\bf 15} (1982) 161.\
\bibitem{salzner} U. Salzner, P.G. Pickup, R.A. Poirier and J.B. Lagowski,
{\it J. Phys. Chem.} {\bf 102} (1998) 2572.\
\bibitem{bakowies1} D. Bakowies and W. Thiel {\it Chem. Phys. Lett.} {\bf 192}
 (1992) 236-242.\
\bibitem{dewar} M.J.S. Dewar, H.S. Rzepa, {\it J. Am. Chem. Soc.} {\bf 100} 
(1978) 784.\
\bibitem{thiel} W. Thiel {\it Tetrahedron} {\bf 44} (1988) 7393.\
\bibitem{lee} Y.-S. Lee, M. Kert\'esz, {\it J. Chem. Phys.} {\bf 88} (1988) 2609.\
\bibitem{bakowies2} D. Bakowies and W. Thiel {\it J. Am. Chem. Soc.} {\bf 113}
(1991) 3704-3714.\
\bibitem{dreizler} R.M. Dreizler and E.K.U. Gross, {\it Density Functional
Theory}, Springler-Verlag, Berlin, 1990.\
\bibitem{salzner2} U. Salzner, J.B. Lagowski, P.G. Pickup and R.A. Poirier,
{\it J. Comput. Chem.} {\bf 18} (1997) 1943.\
\end{references}
\end{document}